# Shape error prediction in 5-axis machining using graph neural networks

Huuk, J.*[a], Denkena B. [a], Dhingra A. [b], Ntoutsi E. [b]

[a] Institute of Production Engineering and Machine Tools (IFW), An der Universität 2, 30182 Garbsen, Germany
RI CODE, University of the Bundeswehr Munich, Carl-Wery-Straße 18, 81739 Munich, Germany

* Corresponding author. Tel.: +49-0511-762-5209; fax: +49-0511-762-5115. E-mail address: huuk@ifw.uni-hannover.de

**Abstract**

This paper presents an innovative method for predicting shape errors in 5-axis machining using graph neural networks. The graph structure is defined with nodes representing workpiece surface points and edges denoting the neighboring relationships. The dataset encompasses data from a material removal simulation, process data, and post-machining quality information. Experimental results show that the presented approach can generalize the shape error prediction for the investigated workpiece geometry. Moreover, by modelling spatial and temporal connections within the workpiece, the approach handles a low number of labels compared to non-graphical methods such as Support Vector Machines.

## 1. Introduction

Machine learning and artificial intelligence techniques have become widely applied in modeling the correlations between machining process features and resulting workpiece quality. The predictive capabilities of these approaches have been applied to diverse problems manufacturers face like reducing production times [1], increasing workpiece quality [2] or avoiding unnecessary quality controls [3]. Commonly used modeling approaches include Support Vector Machines (SVM) [4,5] or Artificial Neural Networks (ANN) [6]. These models operate on the premise that a specific set of features is linked to a quality criterion and if that same set of features is observed again, the recurrence of the same value for the quality-criterion is assumed. However, this approach often disregards time- and location-interdependencies of quality-values. Traditional models perceive machining processes as a stream of isolated feature sets with associated quality values, rather than viewing them as a continuous flow of interdependent features that collectively affect the final product. Modeling these complex connections could be achieved using graph neural networks (GNN), a modeling approach which has not been applied to these kinds of problems before. Additionally, GNNs utilize a semi-supervised learning approach, which shows potential for superior performance in datasets with few labeled examples, compared to supervised methods such as SVMs. Overall, the goal of this paper is to leverage the ability of graphs to model spatial structures and test the prediction capabilities for the use case of predicting shape errors in flank milling.

*1.1. Graph neural networks*

Objects in the real world are often defined in terms of their connections to each other - a set of such objects and the connections between them can naturally be expressed as a graph. In fact, graphs are omnipresent and are all around us - real-world objects such as social networks, recommendation systems, business processes, molecules and roads can be easily thought of as graphs. An example is given in Fig. 1, where an extract from the Munich tube-map is represented as a graph. Here the stations are the nodes and the connections between them the edges. Features considered could be e.g. station capacity, means of accessibility or rush hour intervals.

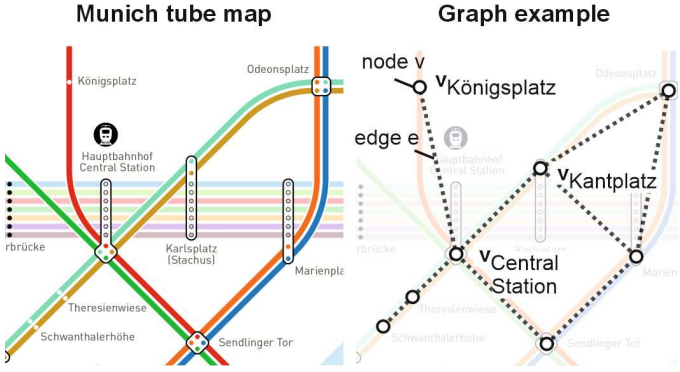

Fig. 1 Graph example created from Munich tube map

A graph is represented by where $V$ is the vertex set with $|V|=n$ nodes $v_i \in V$ and edges $(v_i, v_j)$. The non-negative adjacency matrix is denoted by $\mathbf{A} = [a_{ij}] \in \mathbb{R}^{n \times n}$ (binary or weighted) with a degree matrix $D_{ii} = \sum_j A_{ij}$.

Graph Neural Networks (GNNs) are considered a general approach for defining deep neural networks on graph structured data. GNNs output a vectorial representation, e.g., a real vector, of each node in the input graph by iteratively aggregating features of neighboring nodes and thus updating the feature information of each individual node [7]. The propagation of feature vectors between the graph nodes is performed by a general framework called Message Passing. Even though researchers have been developing GNNs that operate on graphs for over a decade and many variants of GNN architectures have been proposed, the core idea of exchanging information between neighborhood nodes still remains the same. In each iteration, an updated feature vector is computed that encodes the information that each node accumulated through these messages.

In this work, a GNN-based model was utilized to learn representation vectors for every node which are then used for the downstream task of predicting the shape errors. The assumption is that connected nodes in the graph are likely to share similar target values so the individual node can not only depend on its own features, but also utilize the graph structure to aggregate feature information of connected nodes.

```
Algorithm 1 General GNN Architecture
Input  : Graph: G = (V, E); |V| = n; Input features: {x_v, ∀v ∈ V};
         Neighbourhood function: N : v → 2^V
Output: Vector representations z_v ∀v ∈ V

for t in {1,...,t_max} do
    for v in V do
        // Send Messages
        m_v^(t) ← MESSAGE^(t)(x_v^(t));
    end
    for v in V do
        // Aggregate Messages
        h_v^(t) ← AGGREGATE^(t)(m_u^(t)|u ∈ N_G(v));

        // Update Node Representations
        x_v^(t) ← UPDATE^(t)(x_v^(t-1), h_v^(t));
    end
end
return z_v ← x_v^(t_max), ∀v ∈ V
```

Algorithm 1 provides a more formal definition of the working of a general Message-Passing Graph Neural Network (MPGNN) architecture in pseudocode. A GNN takes a graph $G = (V, E)$ and input features $\mathbf{x}_v, \forall v \in V$ and produces a mapping from the input node features to learned representation vectors $\mathbf{z}_v$ for each node. The network allows the nodes of $G$ to exchange information for $t_{max} \in \mathbb{N}$ iterations.

The process of propagation and aggregation of node features throughout the graph is done in three main steps:

- **Generate messages:** A trainable function MESSAGE(·) maps the node features $\mathbf{x}_v, \forall v \in V$ at each iteration $t$ to a message $\mathbf{m}_v$ that is then sent to the nodes neighbors.
- **Aggregation of messages:** A non-trainable function AGGREGATE(·) aggregates all received messages from a node's neighborhood to a single vector. Possible pooling functions are element-wise max-, sum- or mean-pooling.
- **Updating features:** A trainable function UPDATE(·) updates the features of each node based on the received messages and the latest features.

After the maximum number of iterations $t_{max}$ the training of the GNN is complete and the learned representation vectors are obtained for every node. These vectors can then be used for the downstream task of predicting the shape error based on the input features.

*1.2. Semi supervised learning using Graph Convolutional Networks*

The problem considered in this work is semi-supervised regression on graphs, meaning there is only a small subset of nodes containing shape error values. Given a graph $G = (V, E, X)$, where $\mathbf{X}^{(0)} = [\mathbf{x}_1, \mathbf{x}_2, ..., \mathbf{x}_n]^T \in \mathbb{R}^{n \times d}$ is the feature matrix, and $\mathbf{x}_v^{(0)} \in \mathbb{R}^d$ is the $d$-dimensional feature vector of vertex $v$. The set of vertices $V$ contains a subset of labeled vertices $V_l$, the goal is to predict the labels for the remaining unlabeled vertices $V_u$.

One of the more popular variants of GNNs is the Graph Convolutional Network (GCN) which generalizes convolutions in a non-Euclidean domain [8]. The working principle behind the GCN for semi-supervised learning is that the model integrates both the structural information of the graph and its node features during convolution. The features of the unlabeled nodes are then learned through multiple iterations of the message passing scheme in which the features of the labeled nodes are propagated to the neighboring unlabeled nodes [9]. The GCN model was first utilized to perform a semi-supervised node classification task using the following layer-wise propagation rule [10]:

$$X^{(t+1)} = \sigma(\tilde{D}^{-1/2} \tilde{A} \tilde{D}^{-1/2} X^{(t)} W^{(t)}) \quad (1)$$

where $t = \{0, 1, ..., t_{max}\}$, $\sigma(\cdot)$ denotes an activation function, such as ReLU$(\cdot) = \max(0, \cdot)$, $\tilde{A} = A + I$, $\tilde{D} = \sum_j \tilde{A}_{ij}$, $\mathbf{W}^{(t)} \in \mathbb{R}^{(d_t \times d_{t+1})}$ and $\mathbf{X}^{(t+1)} \in \mathbb{R}^{(n \times d_{t+1})}$ denotes the output of activations after $t$ iterations of message passing.

In this work the final perceptron layer for the semi-supervised regression is defined as:

$$Z^{(t_{max})} = \mathrm{MLP}(\sigma(\tilde{D}^{-1/2}\tilde{A}\tilde{D}^{-1/2}X^{(t_{max})}W^{(t_{max})}), W_{\mathrm{MLP}}^{(t_{max})}) \quad (2)$$

where $\mathrm{MLP}(\cdot)$ is a fully connected neural network with weight parameters $W_{\mathrm{MLP}}^{(t_{max})} \in \mathbb{R}^{m \times 1}$, $\mathbf{X}^{(t_{max})} \in \mathbb{R}^{n \times d_{(t_{max}-1)}}$, and $\mathbf{Z}^{(t_{max})} = [z_1, z_2, \ldots, z_n]^T \in \mathbb{R}^{n \times 1}$ denotes the predictions for all data $X$ generated from the multilayer GCN model with weight parameters $\mathbf{W}^{(t_{max})} \in \mathbb{R}^{(d_{(t_{max}-1)} \times m)}$.

The optimal weight matrices $\mathbf{W}_{\mathrm{GCN}} = \{\mathbf{W}^{(0)}, \mathbf{W}^{(1)}, \ldots, \mathbf{W}^{(t_{max})}\}$ and $\mathbf{W}_{\mathrm{MLP}}$ are trained by minimizing the L2 loss or the Mean Squared Error as,

$$\mathcal{L}_{Semi-GCN} = \frac{1}{|V_l|} \sum_{v \in V_l} (Y_v - Z_v)^2 \quad (3)$$

Where $Y_v$ is the ground truth label and $Z_v$ is the predicted value.

A fundamental aspect of GCN involves representing the feature information X as a graph $G = (A, X)$. In certain cases, the graph structure is already known for naturally occurring graphs such as social networks, recommender systems, molecules, etc., which makes it easier to directly utilize a GNN model for semi-supervised learning tasks.

However, in many cases, datasets being investigated are not inherently structured as a graph (e.g., tabular data, textual data, etc.) but require preprocessing to define the connections between nodes. A common approach involves manually constructing graphs based on human expertise e.g. by employing a k-nearest neighbor strategy to establish connections [11,12]. This shows that finding the optimal graph structure is a challenge and graphs constructed based on domain knowledge often require significant trial and error to accurately capture the inherent graph structure of the data.

## 2. Procedure

### 2.1. Dataset generation for shape error prediction

The data used to evaluate the suitability of the GNN to model shape errors from flank milling processes consists of three data streams: Quality data, process data and simulation data. The quality data represents the actual measured shape errors at the workpiece. Machining was conducted on the 5-axis milling center DMG MillTap 700 using a solid carbide tool with d = 6 mm and four teeth for the finishing process. The shape error was tactilely acquired by using the machines built-in Blum TC52 touch probe. Process data is recorded at 150-200Hz using the software library AGLink and consists of 43 values e.g. actual axis positions $x, y, z, a, c$, currents $I_{x,y,z,a,c}$, actual feed rates $v_{f_{x,y,z,a,c}}$ and acceleration values $a_{x,y,z,a,c}$. Simulation data provides six different engagement conditions e.g. material removal rate $Q_w$, tool engagement depth $t_{height}$, and chip cross section $A$. To generate the simulation data the technological material removal simulation IFW CutS [13] is employed. The simulation is performed after the machining of the part using the process data to calculate the engagement between tool and workpiece at discrete time steps. In IFW CutS a 3D Cartesian Dexel-model is utilized to discretize the workpiece, where each Dexel start- or endpoint will become a node in the GNN. During a single time step in the simulation anything between zero to multiple hundred Dexel are modified, depending on the discretization density applied and the current engagement between tool and workpiece. All modified Dexel are subsequently assigned the resulting simulation data for that time step. During the export of the final workpiece geometry, the simulation data is merged with the quality and process data, resulting in datasets linking the time at which Dexel start- or endpoints were cut, their positions in the 3D space and the corresponding features. All datasets are published under the creative common license with a more detailed annotation [14].

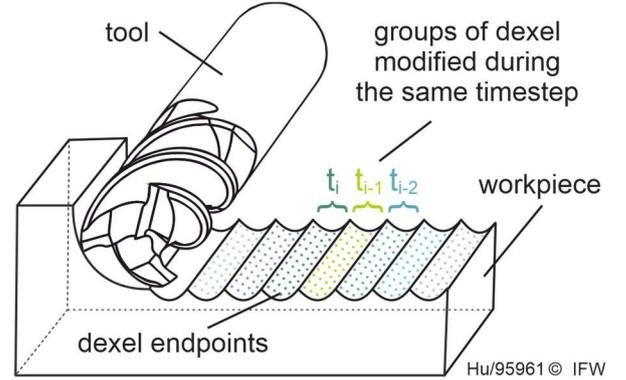

Fig. 2 Dexel endpoints grouped by time step

### 2.2. Experimental set up and trial plan

For handcrafted graphs, where the nodes are not naturally connected, the performance depends on the chosen method of connecting the nodes. For the prediction of the shape error, all Dexel start- and endpoints on the surface generated during the finishing operation of the milling process are regarded as nodes. Two strategies of node connections were explored: k-nearest-neighbors (kNN) and forming connections determined by the time step (Table 1).

Table 1 Trial plan

| Parameters | Values | | | | |
|---|---|---|---|---|---|
| **Filtering of nodes** | | | | | |
| Filter percentage [%] | 0 | 10 | 50 | 90 | 99 |
| Minimum number of nodes | 5 | 5 | 5 | 5 | 1 |
| Ratio labeled / unlabeled nodes | 0.065 | 0.069 | 0.130 | 0.217 | 0.510 |
| **Connection strategy** | | | | | |
| kNN (number of neighbors) | K = {3, 4, … ,8} | | | | |
| Temporal (number of time steps considered) | T = {1, 2, … ,8} | | | | |

When nodes are connected via the kNN strategy, the $n$ geometrically closest nodes are connected for each node. When connecting via the time steps, each node is connected to a single node from each of the $n$ preceding time steps. This means if the past three time steps are connected, a node from the time step $t_i$ is connected to a single node from $t_{i-1}$, $t_{i-2}$ and $t_{i-3}$. Further, the density of the dexelcore was modified to explore the predictive performance for more sparsely populated graphs. Here the nodes were filtered while preserving a minimum amount of nodes per time step and all labelled nodes. Thus, a filter of 50% means that for each time step 50% of the modified nodes are randomly dropped, while ensuring that all labelled nodes and a minimum of 5 nodes are preserved.

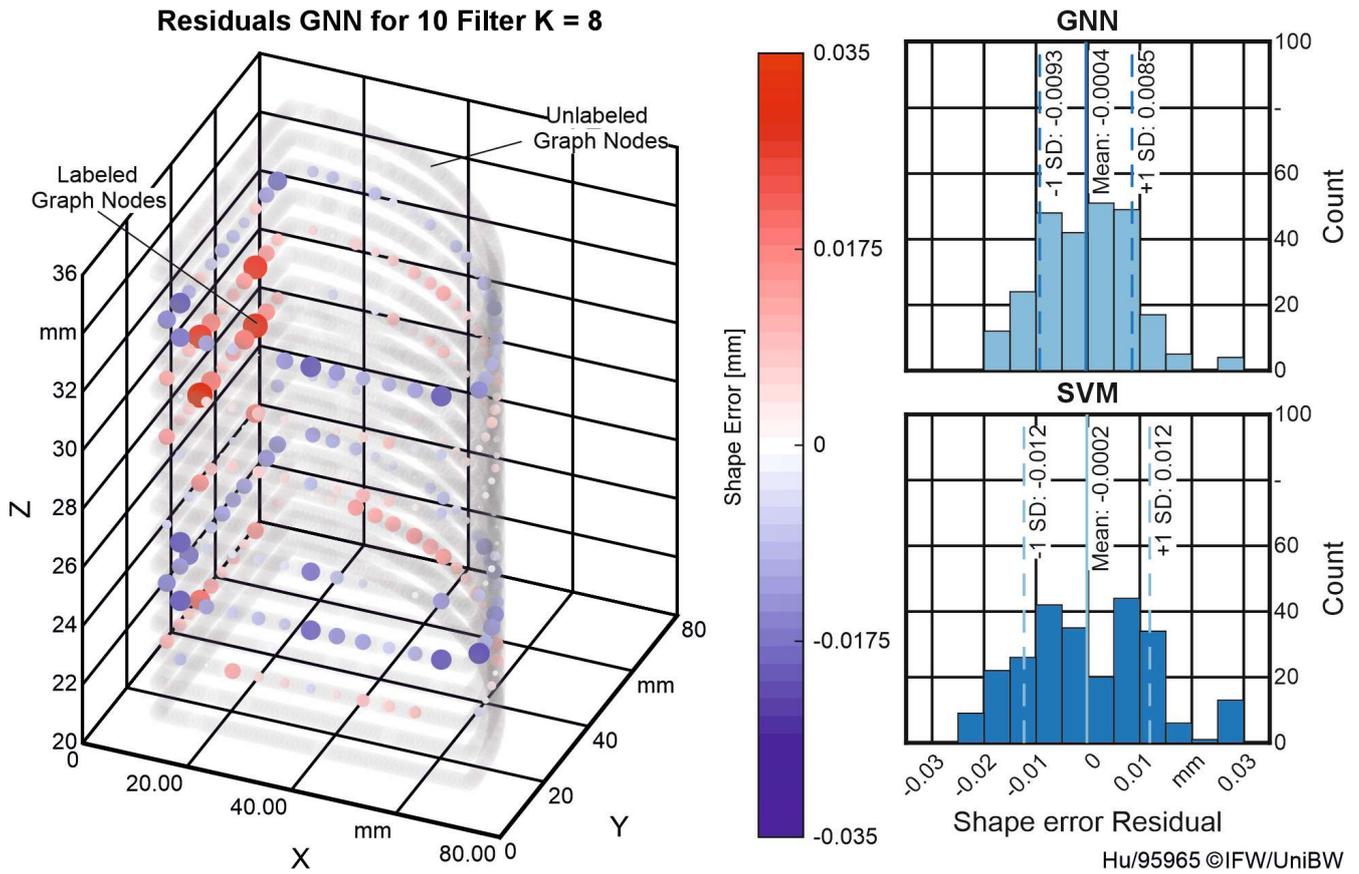

Fig. 3 Left: Plot of all nodes in 3D grid, unlabeled nodes in grey, labeled nodes coloured according to their shape error residual. Right: Distribution of shape error residuals for GNN and SVM predictions.

A three-layer GCN coupled with a fully-connected layer (MLP) was trained as described in section 1.1. The model was trained for a maximum of 500 epochs using the Adam optimizer [15] with a learning rate of 0.005, a weight decay of $5 \cdot 10^{-4}$ and a dropout rate of 0.6. A hidden layer size of 20, a dropout rate of 0.6, and randomly initialized the weights and (row-) normalized the input feature vectors was used. The GCN model is compared against a Support Vector Regressor (SVR) as the baseline method. A grid search for optimizing hyperparameters of the SVR was performed. Further, the performance of the GCN and SVR were compared using k-fold cross-validation.

### 2.3. Evaluation of predictive performance

After training the GNN models and using them to perform predictions for all nodes, the prediction values for the shape error were compared to the original data set.

In Fig. 3 the residuals are plotted for the simulation trial performed with a 10% node-filter and connecting the eight nearest neighbors, the histograms show the distribution of residuals for the GNN and SVM. A positive residual means the model predicted a lower value than was measured. Comparing the predictive performance between the three different faces of the workpiece, there is no noticeable dependency between the predictive performance and node-positioning on the workpiece. The residuals scatter around a mean value of approximately zero. This indicates that the model generalizes well to unseen data points despite having been trained on only few labeled data points.

### 2.4. Insights for Graph construction

Comparing the mean absolute error (MAE) for different numbers of connected nearest neighbors' shows a trend that connecting more neighbors decreases the MAE (Fig. 4).

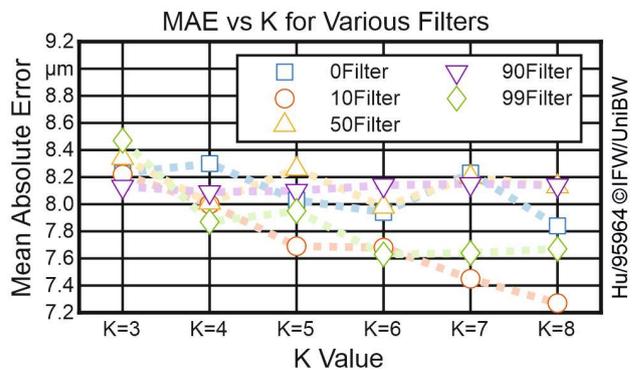

Fig. 4 Effect of the increase in geometrical neighbors connected

This indicates that GNNs on more densely connected graphs tend to perform better. The highest number of neighbors investigated (K=8) represents a fully connected node in a 2D grid structure. The results of performing predictions for various filters show no distinct trend (Fig. 5). Employing a more densely connected graph, however results in a lower MAE for most filters, with the 90% filter showing almost identical MAE-values for all K-values. While filtering the data does not increase the predictive performance, this analysis shows that the employed GNN consistently provides good results independent from the density of nodes in the graph or the ratio between labelled and unlabeled nodes.

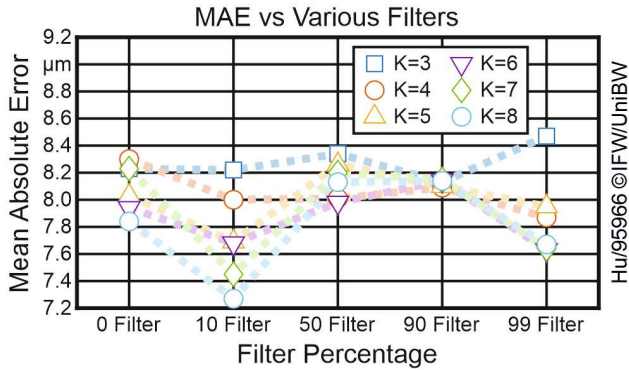

Fig. 5 Effect of increasing filtering percentages

Fig. 6 shows the model's predictive accuracy for connecting different corridors of time steps. A value of T=3 corresponds to connecting each node to one node from each of the three previous time steps. However, because this is done for every node and the connections are bidirectional it leads to many nodes also being connected to nodes from the following three time steps. This setup allows the model to leverage temporal information, aiming to enhance prediction accuracy by learning time-related impacts on the shape errors.

A notable improvement in the model's performance is observed specifically up until T=3. However, further increasing the number of connected time steps beyond T=3 does not appear to yield additional increases in performance. When comparing the performance of models on graphs built with the temporal connection strategy against those using the kNN connection strategy, it is evident that the kNN strategy achieves lower minimum MAE values. However, the improvement observed up until T=3 suggests that a hybrid approach combining the temporal and kNN connection strategies might be beneficial. The results indicate an enhanced predictive performance with a greater number of connected spatial neighbors, suggesting that the GNN is capable of managing this increased level of connectivity without causing over smoothing in the predictions.

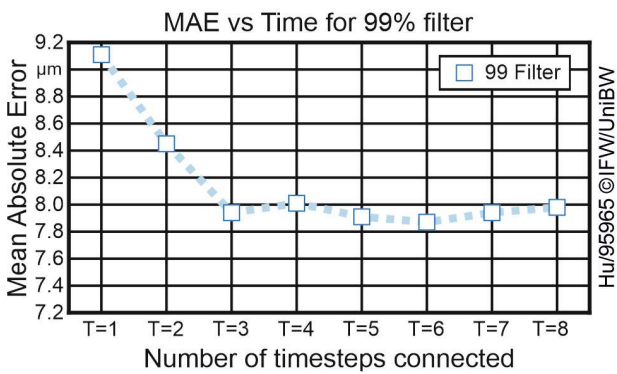

Fig. 6 Effect of the increase in connected time steps

The results of the simulation trials showcase that there currently is no straight forward process to find an optimal configuration for a handcrafted graph. While an increase in the number of neighbors using the kNN connection strategy seems beneficial for the predictive performance, it is also evident that choosing the "right" neighbors seems more relevant than the number. Further investigations on hybrid connection strategies seems promising.

### 2.5. Transferability

To evaluate the transferability of learned correlations between feature vectors and shape errors, two GNN models were tested on a workpiece with a more complex geometry, distinct from the training set. These models, characterized by settings of 10% filter, K=8, and 99% filter, K=8, were applied to predict shape errors on this new workpiece, which features a more curved contour and a chamfer machined using 5-axis milling strategies. The trained models were used to perform predictions for the transfer geometry applying a 0% filter and the kNN-connection strategy with K=8. The MAE was assessed separately for the side surfaces and for the chamfer, with the results presented in Table 2.

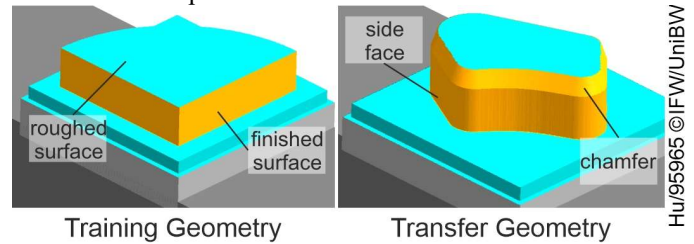

Fig. 7 Simulation result from IFW CutS for workpiece geometry used for training and transfer

Table 2 Comparison of MAE for transfer and training workpiece geometries

| Workpiece | MAE [µm] | | |
| --- | --- | --- | --- |
| | Transfer | | Training |
| Model | Side | Chamfer | Overall |
| GCN (10% filter, K=8) | 0.0241 | 0.0281 | 0.00727 |
| GCN (99% filter, K=8) | 0.0299 | 0.0515 | 0.00767 |
| SVR | 0.0176 | 0.0147 | 0.0124 |

It can be observed that the GCN models trained on the workpiece graph of different densities fail to perform well on unseen nodes of a different workpiece geometry. While semi-supervised learning using GCN has seen much success under a transductive setting, where only labels of the unseen nodes in the training data set need to be inferred, the transfer creates a setting where labels for unseen nodes need to be inferred. This is still a relatively unexplored research direction and the results show that there is much potential left. A potential approach for generating accurate predictions to test the transferability of the GCN models would be exploring the use of transfer learning with a pre-trained GCN from one workpiece geometry and then fine-tuning it on the graph constructed for the target workpiece geometry before testing [16].

*2.6. Comparison between SVM and GNN performance*

In comparing the performance of the SVM and the GNN, it's notable that the SVM produces a MAE of 0.0124 mm across all filters. Thus, the GNN modelling approach consistently outperforms the optimized SVM baseline approach. Additionally, as illustrated in Figure 2, the distribution of residuals for the GNN models is more tightly clustered around zero, which is generally seen as a preferable outcome. One reason for the higher MAE values produced by the SVM might be attributed to SVMs typically requiring feature selection to eliminate correlated features, as they are not optimized to handle a large number of features.

The transferability tests discussed in section 2.5, the SVM produces a lower MAE than both tested GCNs. However, the SVM's predictions are very uniformly distributed within a ±0.5 µm range. While these predictions align better with the measured shape errors, the uniform distribution is not found in the shape error data.

## 3. Conclusion and Outlook

This paper demonstrated the potential of GNNs for predicting shape errors in milling, showcasing promising results with MAEs between 7.2 and 9.2 µm. The findings highlight the capability of GNNs to model complex interdependencies inherent in machining processes. Notably, GNNs perform well in scenarios with low percentage of labels and can leverage the structural and temporal connections of data points to improve prediction accuracy.

Despite these advancements, the transfer to a different workpiece geometry, showed varied performance for the GNN investigate in this setting. This indicates a need for further research into model generalization and suggests that transfer learning could be a viable path forward. By fine-tuning pre-trained GNNs on new geometries, their applicability and robustness for practical machining tasks may be enhanced. Accurately predicted shape errors can consequently be used to optimize process planning approaches to compensate the shape errors for manufactured goods.

**Acknowledgements**

The authors would like to thank the German Research Foundation (DFG) for its funding of the research project "Hephaestus" (project number: 424298653).